# Circular photogalvanic effects in topological insulator/ferromagnet hybrid structures

[1]T. Schumann, [1]T. Kleinke, [2]L. Braun, [1]N. Meyer, [3]G. Mussler, [3]J. Kampmeier, [3]D. Grützmacher, [4]E. Schmoranzerova , [5]K. Olejník, [5]H. Reichlová, [6]C. Heiliger, [1]C. Denker, [1]J. Walowski, [2]T. Kampfrath, [1]M. Münzenberg

1. Institut für Physik, Universität Greifswald, Felix-Hausdorff-Straße 6, 17489 Greifswald, Germany
2. Department of Physical Chemistry, Fritz Haber Institute, Faradayweg 4-6, 14195 Berlin, Germany
3. Peter Grünberg Institut (PGI-9) and Jülich-Aachen Research Alliance (JARA-FIT), Forschungszentrum Jülich, 52425 Jülich, Germany
4. Faculty of Mathematics and Physics, Charles University, Ke Karlovu 3, 121 16 Prague, Czech Republic
5. Institute of Physics ASCR v.v.i., Cukrovarnická 10, 162 53 Prague, Czech Republic
6. Justus Liebig Universität, Gießen, Germany

Abstract

We study laser driven spin-current effects at ferromagnet/topological-insulator interfaces by two complementary experimental approaches. The DC photocurrent is studied in ferromagnet/topological-insulator bilayers with high spatial resolution. Dynamic interface currents are explored via the emission of terahertz radiation emitted by these currents with high temporal resolution. From our experiments, we reveal a lateral and dynamic interaction of the ferromagnet and the topological insulator interface.

I. Introduction

A topological insulator is a bulk electrical insulator with a two-dimensional metallic surface. These films are of large interest because of spin-momentum locking of the surface states which allows one to drive spin dynamics at ferromagnet/topological-insulator (FM/TI) interfaces [1, 2]. The combination of a ferromagnet and a topological insulator enables to manipulate the magnetization of the ferromagnet by spin-orbit torques.
Recent results have shown that a surprisingly large conversion factor of a charge current into a spin currents seem possible [3]. The conversion efficiency is given by the spin Hall (SH) angle which is defined by $\Theta_{\text{SH}} = \vec{j}_{spin}/\vec{j}_{charge}$ where J$_{spin}$ and J$_{charge}$ represent the corresponding spin- and charge current densities. While normal metals show efficiencies up to a maximum of 0.3, the record value for FM/TI interfaces suggests that $\Theta_{\text{SH}}$ can amount up to 100 and even more, making these interfaces very interesting for charge-to-spin conversion in spintronic devices. Current theoretical investigations of these interfaces suggest that the efficiency can be only explained when the interface channel plays a significant role, and the bulk transport channels are suppressed [4]. Therefore, granular films of a topological insulator could be more efficient for spin-orbit-torque-based switching of the ferromagnet's magnetization. Second, a ferromagnet can also be used to confine and control the surface states [5]. In FM/TI films, the edges show quantized conductance, and their conduction can be controlled by the magnetic state. The direction of the surface state is, owing to spin-momentum locking, spin polarized and unidirectional [6] [7] [8] [9]. It has been suggested to use domain walls to shape complex devices of a topological insulator based spintronics [10, 11]. Third, the use of all-optical control of the transient magnetization of topological insulators has been demonstrated [12]. Exploiting the selection rules and the helicity of the light, spin currents can be driven all-optically by the circular photogalvanic effect. In the following, we aim to combine all-optical excitation of the surface and interface currents in ferromagnet/topological-insulator double layers $(Bi_{1-x}Sb_x)_2Te_3$/CoFeB and focus on the composition of around Sb=50%. We excite the ferromagnet/topological-insulator sandwich structure optically and probe the photocurrent response of the electrons for different helicities of the exciting laser beam. The photogalvanic effects are detected in two ways. We measure them with standard contacts in lateral devices and then, we will detect the transient interface current by the terahertz (THz) emission spectroscopy in a contact-free manner. The absence of contacts allows us to also measure the temporal evolution of the currents and at buried interfaces.

## II. Experiment

The topological-insulator films were prepared by molecular-beam epitaxy [13] and characterized by structural characterization: x-ray diffraction measurements, Raman scattering, angle-resolved photoemission spectroscopy (ARPES) and transport experiments to extract their Fermi level and carrier density for an Sb content ranging from x = 0 to 1 [14]. Strain free films are realized by van der Waals epitaxy on a 70 nm thick Si-(111) surface with an insulating layer of oxidized silicon of 300 nm thickness top layer on the Si wafer. The chosen composition of $(Bi_{57}Sb_{43})_2Te_3$ films is intrinsic or has very little n-doping. It is found that it is as close as possible to the Dirac point. The carrier density is below $3 \times 10^{13}$ cm$^{-2}$.

To prepare the FM/TI bilayers, we used 15 to 160 nm thick films that undergo surface cleaning and in-situ Ar-ion etching. The surface layer composition is checked by its XPS spectra during the whole process. The few-nanometer thick oxidized surface is cleaned until the pure Te peak appears in the XPS spectra and the oxide peak is removed. The spectra for Te, as well as for $TeO_2$ is shown in Fig. 1a. A careful 5 minute long in-situ annealing at 250°C heals the surface defects due to etching, but does not change the surface stoichiometry significantly, as confirmed by ARPES. A successful reconstruction of the Dirac cones and surface states was demonstrated [15] and is presented in Fig. 1c. Subsequently, a 3 nm thick $Co_{0.2}Fe_{0.6}B_{0.2}$ layer was deposited by plasma sputtering in the same UHV system at a rate of 0.4 Å/s at a base pressure of $10^{-9}$ mbar. Since the in-situ Ar-ion etching beam is focused and scanned on the substrate, it allows to prepare wedges of the $(Bi_{0.57}Sb_{0.43})_2Te_3$ film. Finally, a 2.5 nm MgO capping was added for protection.

### A. Photocurrent microscopy

The device layout is presented in Fig. 2. It consists of a large square (area 200×200 µm²) of plain topological insulator film. A smaller square (40×40 µm²) containing the ferromagnetic layer is patterned in the middle. The whole device is contacted by four contacts along the sides and four at the edges, thereby allowing for a measurement of the photo- and thermovoltages along all directions and. In addition, by using the contacts, a current can be driven through the structure. The device was prepared by three-step electron-beam lithography. In the first and the second step, the squares were defined by writing with the electron beam, and a hard etching mask of Al (50 nm) was deposited using lift-off technique. The individual structures were defined by Ar+ ion milling, and the mask was removed after each step. Finally, large Au (90 nm) contacts were defined by optical lithography and lift-off.

The experiment schematic for the DC photocurrent measurements is presented in Fig. 2a, while Fig. 2b shows a large-scale optical reflectivity map of our device. A Gaussian laser beam (wavelength 785 nm, 1mW) was focused down to a waist of 3 µm scanned over the sample. For each point of the map, the photovoltage $V_{photo}$ is measured for different polarization of the light. By rotating the quarter-wave-plate angle $\theta$ before the sample by 360°, the polarization is changed from linearly to right-handed circularly to linearly and left-handed circularly polarized within one circle of 360°. A modulation frequency of 77 Hz was applied by square-type intensity modulation of the laser beam. The resulting signal was analyzed using the relationship

$$V_{photo} = C \sin(2\theta) + L_1 \sin(4\theta) + L_2 \cos(4\theta) + D, \tag{1}$$

taken from Ganichev et al. [16] [17]. By fitting, we obtain the parameters D, $L_1$, $L_2$ and C at each point of the laser spot. The coefficient D corresponds to the thermovoltage, while C quantifies the strength of the circular photogalvanic voltage driven by right and left circularly photons in the sample. $L_1$ and $L_2$ correspond to the real part and imaginary part of the linear photogalvanic voltage, in phase and out of phase by $\pi/2$ with respect to $\theta = 0$.

The resulting map of D is shown along with an optical reflectivity micrograph in Fig. 3a. To interpret the signals, we emphasize that the laser scanning generates a thermal gradient and a (helicity) dependent photocurrent. At each sample position, the laser generates a different thermal profile and thermovoltage. For laser excitations close to the Au contact pad we find large values of D, which indicate an inhomogeneous thermovoltage driven at the Au/SiO/Si(111) interface from the semiconductor/metal contact. A detailed close-up of the 2D map and averaged line profiles above and at the side are shown (top and right side panel). When moving the laser spot perpendicular to the contacts no slope indicate constant thermovoltages (right panel). When moving the laser spot illumination parallel to the contacts (top panel), we observe a sign change with different slopes on the topological insulator slightly steeper as on the ferromagnetic element with steps at the edges.

The reflectivity signal is determined by the reflected laser light at the same time. This is helpful to quantify the local absorption. The reflectivity map shows that the ferromagnetic square element has a higher reflectance, caused by the ferromagnetic layer and residual remaining Al etch-layer on top. Therefore, we expect a smaller thermoelectric signal here. This is reflected in the behavior of D, the thermovoltage levels off on the ferromagnet element and shows a plateau. On the topological insulator the thermovoltages show a larger slope reverse the sign, while the laser spots move from one contact to the other, demonstrating that we can successfully create controlled temperature gradients by the laser heating in between the contacts in the topological insulator film.

Next, we want to study the area at around the topological insulator and the interface to the structured FM/TI element. The resulting map of C, sensitive to the circular chirality of the photogalvanic voltages extracted is shown in Fig. 4. A schematic representation how this spin accumulation is detected by the spin-galvanic is given in Fig. 4. The optical excitation in the area of the accumulated spins results in a small, detectable circular photo-galvanic current, resulting in a voltage in the µV range. While the size of the detected voltages is small, the absolute change in signal is quite large, which points indirectly to a high degree of spin polarization. Again, a detailed close-up of the 2D map and averaged line profiles above and at the side are shown (top and right side panel). The C parameter, shows a completely different behavior. This shows that it allows to discriminate between thermal effects, and effects driven by the circular photogalvanic effect. Both effects can be well separated.

In the area of the free topological insulator film, the value of C stays constant. In the proximity of the edges of the ferromagnetic structure within a short distance below a few micrometers, significant changes are found. The circular photo-galvanic effect for right and left shows two peaks for the vertical average over a line section. One positive and one negative peak is found. The negative and positive peak occurs perpendicular to the thermocurrent flow, e.g. perpendicular to the thermal gradient, next to the FM/TI element edges. They have the signature of a spin-Hall effect, arising from a spin-to-charge conversion by spin-orbit scattering [18], driven by the thermal gradient (spin-Nernst effect). In addition, we used a +31mT or -31mT magnetic field applied in-plane and along the temperature gradient as a control of the magnetization direction. While the right panel remains unchanged, we observe now a splitting of the circular photogalvanic voltage parallel to the thermocurrent flow, e.g. along the thermal gradient. This splitting can be controlled by the field direction. The voltage signal switches with the +31mT or -31mT magnetic field rotating the magnetization of the TI/FM element. The signature of the ferromagnet topological insulator interaction can only be understood by a significant modification of the surface states. On the right side of Fig. 4, a schematic representation is given of the band structure of the topological insulators surface state (TSS) is given as inset. Schematically we show magnetic interfaces or dopants interact with the Tis surface bands opening a gap, and shifting the bands along the k-axis. The spin-currents arising from the spin-Nernst effect appear to be blocked at the edges of the ferromagnetic element on top of the topological insulator film. Our spatial mapping allows to determine the length scale of the accumulation at the interface of two to three micrometer. We conclude that in this case the ferromagnetic structure acts in a similar way as if the topological insulator would have finite edges. This results in a spin-accumulation at the edges of the ferromagnetic element. Two main effects have been discussed at these interfaces, by calculation of the interface band structure in presence of the ferromagnetic layer on top of the TSS by density functional theory and by developing model Hamiltonians of the effect of the exchange interaction. A spin Hamiltonian developed by Kim et al. [27] is used in the following,

$$H_{TSS} = v_F(-i\hbar \nabla \times z) \cdot \sigma - J(m_x \sigma_x) + m_y \sigma_y) - J_\perp(m_z \sigma_z), \qquad (2)$$

where $v_F$ is the Fermi velocity, $\sigma$ is the spin operator (Pauli spin matrices) of the conduction electrons, $m_i$ denotes the magnetic moment of the magnet, and J is a coupling parameter. This results in the following modification of the band structure:

$$E_\pm = \pm\sqrt{(p_x - Jm_y)^2 + (p_y - Jm_x)^2 + J(p_x - Jm_y)^2}, \quad p = \hbar v_F k \qquad (3)$$

In this equation, we observe two effects: (i) a band splitting at the Dirac crossing point, proportional to J and (ii) a lateral shift of momentum depending on the direction of **M**. We have in our case a magnetization that is in-plane along ±x direction. We observe two different effects along the x and y direction of the edge of the ferromagnet. A band splitting could result in a channel that blocks the spin-transport and corresponds to the finding of a spin accumulation that we observed located at the edge of topological insulator films with finite edges along the y direction perpendicular to **M**. This effect will not be affected by the magnetization's sign but only by its absolute value. In the case of the momentum shift of **k**, we expect a direction magnetization **M** dependent component. In the configuration that the effect of the momentum shift dominates, this will result in and increase and decrease for the two spin directions respectively and thus a spin accumulation that will change its sigh as we observe in the cross section of the photogalvanic signal along the x direction indicating a splitting spin occupation before and after the ferromagnetic structure that changes its sign. It demonstrates the capability to switch the spin-occupation in the topological insulator via the magnetization of a micron or nanostuctured ferromagnetic element in proximity to a topological insulator.

## B. THz emission spectroscopy

The question remains, if below the ferromagnet, in proximity to the topological insulator to the ferromagnetic film, currents are flowing, or if the spin-currents below the ferromagnet are blocked completely as one might suspect at first glance. To detect an optically driven transfer of spin polarization from the ferromagnet to the topological insulator, we performed photocurrent experiments with ultrashort optical pulses (duration ~10 fs, wavelength 800 nm) as schematically shown in Fig. 5. Any optically induced spin current flowing from the ferromagnet to the topological insulator [28,29] will be converted into an in-plane charge current by spin-to-charge-current conversion (S2C). As the resulting current burst is ultrashort, it will emit electromagnetic radiation with frequencies extending into the terahertz (THz) range. The transient THz field can be measured by electro-optic sampling.

Typical electro-optic waveforms of emitted THz pulses are shown in Fig. 5a. When we reverse the direction of in-plane magnetization ***M*** of the FM layer by an external magnetic field, we observe an almost complete reversal of the THz signal. By taking the difference of the two waveforms, we extract the THz signal component linear (odd) in the magnetization. Note that the THz emission signal can also arise from the magnetic-dipole radiation that is emitted by the optically quenched magnetization of the FM layer [33]. For this purpose, we repeated the experiment with a single FM layer and found that the odd-in-***M*** signal component was one order of magnitude smaller than from the TI/FN stack (not shown). This observation strongly suggests that the THz signal from the TI/FN stack indeed arises from optically induced spin transport [28,29] across the FM-TI interface in conjunction with spin-to-charge-current conversion (see Fig. 5).

Our results indicate that the interface between the FM and the TI is "active". A spin current can be injected and converted into a charge current. A comparison to the emission from FM|Pt layers shows, however, that at THz frequencies, our FM/TI stacks are not yet as efficient THz emitters as heavy-metal based stacks such as FM/Pt.

It is not yet clear where and by which microscopic mechanism spin-to-charge-current conversion occurs. Important conversion mechanisms include the inverse spin Hall effect (ISHE) in the bulk of the TI and/or FM and at the interface through the inverse Rashba Edelstein effect (IREE). In an attempt to address this question, we varied the thickness of the TI by using a wedge-like TI film. The resulting THz emission amplitude as a function of the TI thickness is displayed in Fig.5. We observe that the THz signal amplitude saturates for a TI film thickness above 10 nm. This phenomena is also known for CoFeB/Pt layers and can arise from (i) a decreased absorptance of the FM film with decreasing TI thickness and (ii) the back reflection of the spin current at the TI-substrate interface [34].

If scenario (ii) were dominant, the signal drop below a TI thickness of 10 nm would imply a spin relaxation length of 10 nm in the TI. This value is much larger than the values seen for heavy metals (~1 nm), Cu (4 nm) [35] and ferromagnets (~1 nm) [36]. We, therefore, consider scenario (ii) rather unlikely. To test scenario (i), we conducted calculations of the pump-pulse energy deposited in the FM layer by using a transfer-matrix approach and the optical constants of the TI and FM at the pump wavelength. The calculations (not shown) indeed confirm that the drop of the THz signal below a TI thickness of 10 nm can be well explained by the TI-thickness-dependent pump absorptance. Therefore, our thickness-dependent measurements still leave the possibility open that spin-to-charge-current conversion can occur in the bulk of both the FM and the TI layer as well as at the interface in-between.

It will be interesting for the future, in how far concepts of using the FM magnetization direction to confine edge currents inside the topological insulator. As suggested by the Tokura group [7], this apporach would provide access to the shaping of complex spin-current devices in ferromagnet/topological insulator hybrid systems. Indeed, recently, 2D materials have successfully been exploited as spin-filter efficient materials [30-32].

### III. Summary

We presented investigations of the circular photo-galvanic currents for intrinsic doping $(Bi_{1-x}Sb_x)_2Te_3$ with x = 43% CoFeB bilayers patterns of micrometer dimensions that are investigate by scanning optical excitation using circular polarized light and detecting the thermo- and circular photo-galvanic contributions. We show that a detailed separation of contacts and films is necessary. Close to the ferromagnet/topological insulator structure, we observed a spin accumulation that can be compared to the spin accumulation effects observed at a topological insulator boarder. The spin accumulation occurs on a length scale of below 2 to 3 µm. We interpret this signal as a modification of the circular photocurrent spin current by the thermocurrent in the topological insulator structure. In addition, by using THz emission spectroscopy, we demonstrated that the interface is "active" and a significant spin-current flow across the TI-FM interface can be driven by laser excitation. Additional investigations have to follow to clarify whether the spin-to-charge-current conversion predominantly proceeds at the interface or in the bulk. Our observation may provide new methods to structure and control the confinement of the spin-flow in the topological insulator. The spin accumulation can be modified on micrometer length scales. This shows that by engineering the ferromagnet, the lateral spin polarization in the topological insulator can be modified [19] and that we have active interfaces in our devices that allow a control of the spin-current via the magnetization of the top ferromagnetic layer.

Figures:

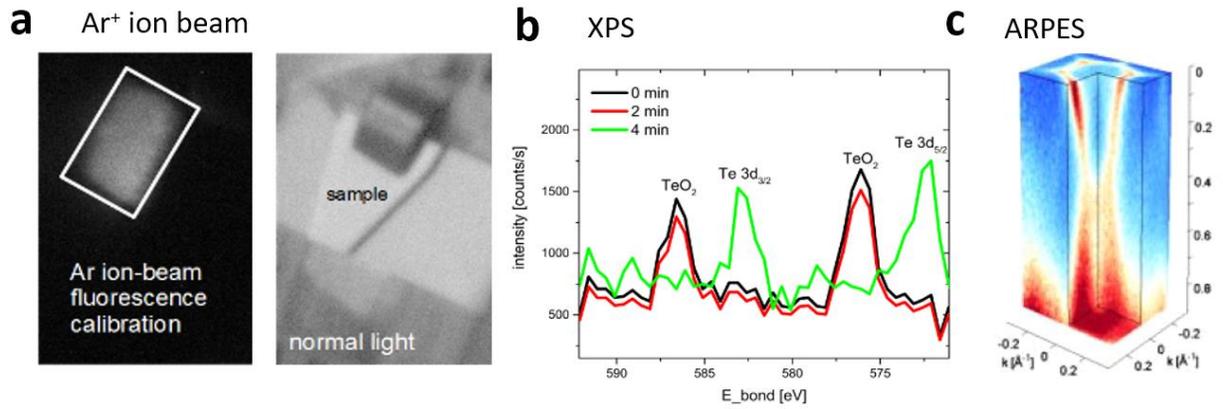

Figure 1: Etching, surface cleaning and characterization of a $(Bi_{0.57}Sb_{0.43})_2Te_3$ film. a) Lateral (in)homogeneity, b) X-ray photoemission spectroscopy (XPS) during surface preparation using Ar+-ion beams and c) ARPES data (APRES data from [15]).

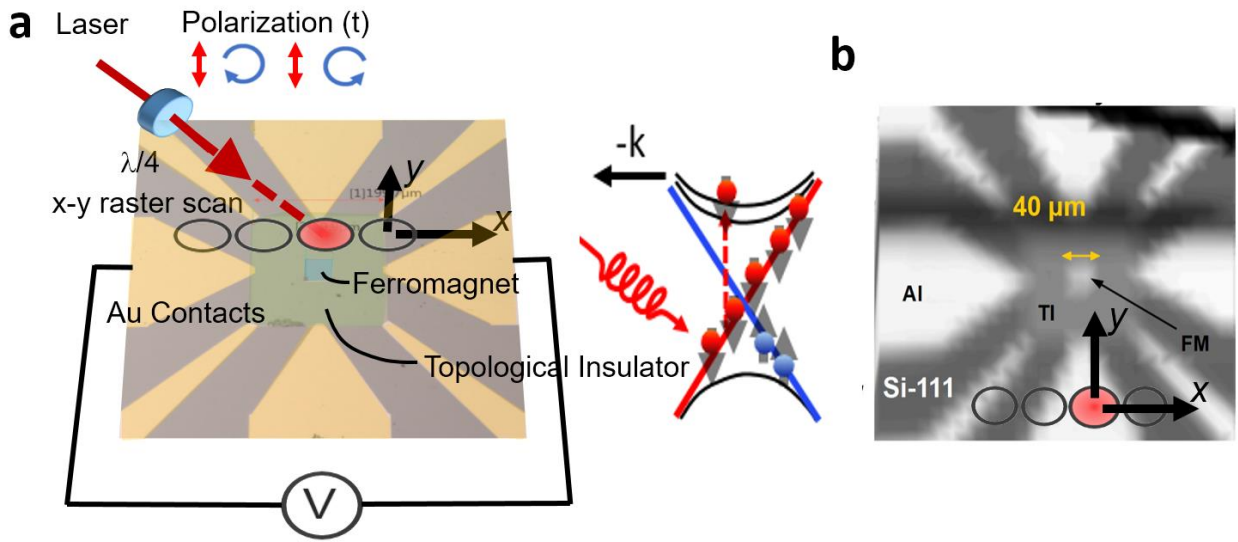

Figure 2: a) Schematic of a ferromagnetic CoFeB element (40 μm × 40 μm) on top of the 200 μm × 200 μm topological insulator $(Bi_{0.57}Sb_{0.43})_2Te_3$ film along with driving laser beam and contacts for photocurrent analysis. By changing the helicity of the laser beam polarization, we can separate various contributions to the photo current at each illuminated point, namely the thermoelectric and light-helicity-dependent photo-galvanic contributions. b) Scanning reflectivity micrograph of the device.

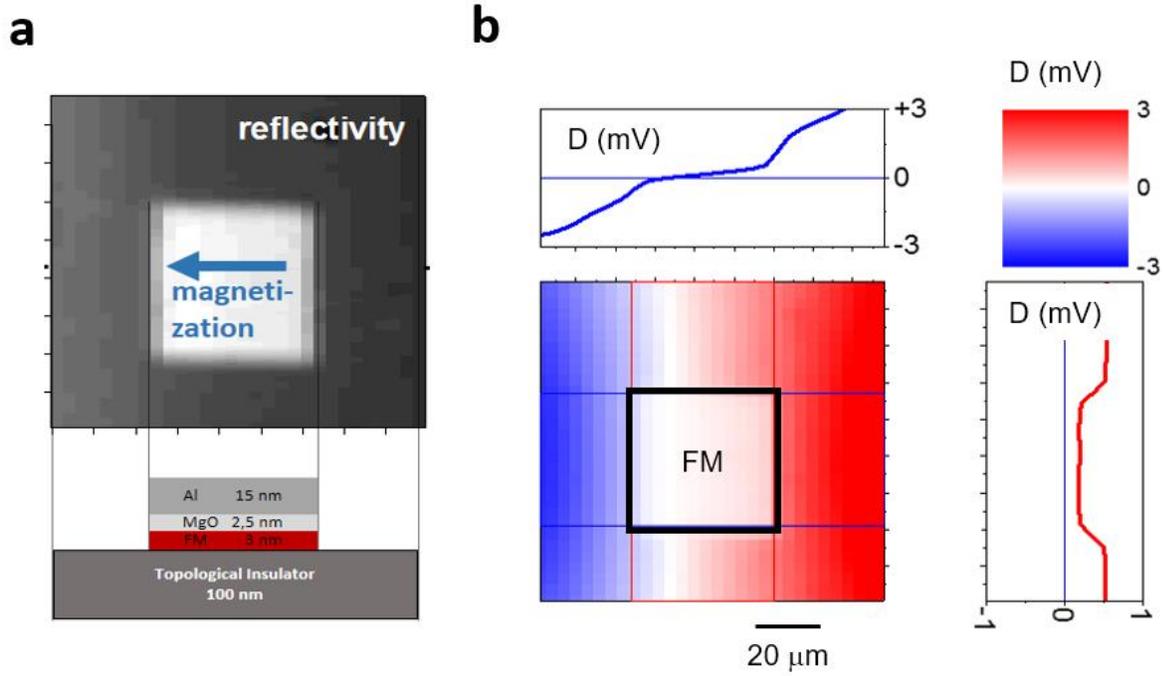

Figure 3: Details of the ferromagnetic (FM) film structure on top of the topological insulator film. a) Optical reflectivity map recorded simultaneously to the b) thermovoltage map. In addition to the color map, the averaged cross sections are shown perpendicular and parallel to the contact pads.

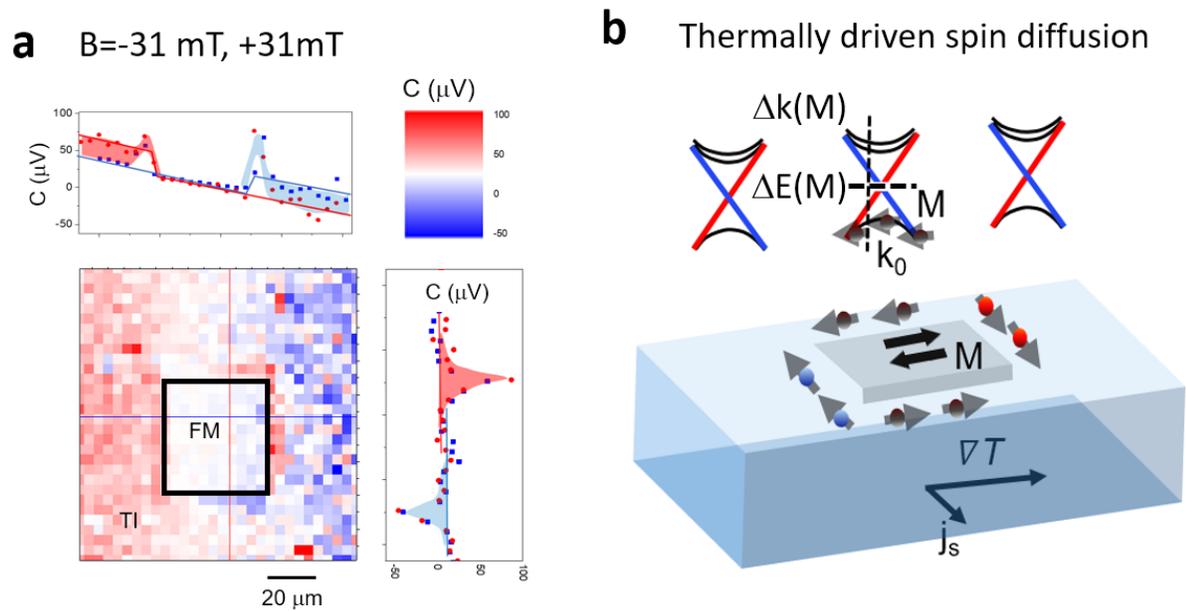

Figure 4: a) Value of the circular photogalvanic effect C. Cross sections are plotted for +-30 mT applied field. b) Schematics of a control of surface current in the topological insulator (FM: ferromagnet, TI: topological insulator). The Dirac cone can be opened, depending on the magnetizations orientation or the shifted. The shades area shows the difference in chirality dependent circular galvanic current for +- B field and the Nernst effect at the edges of the structure respectively.

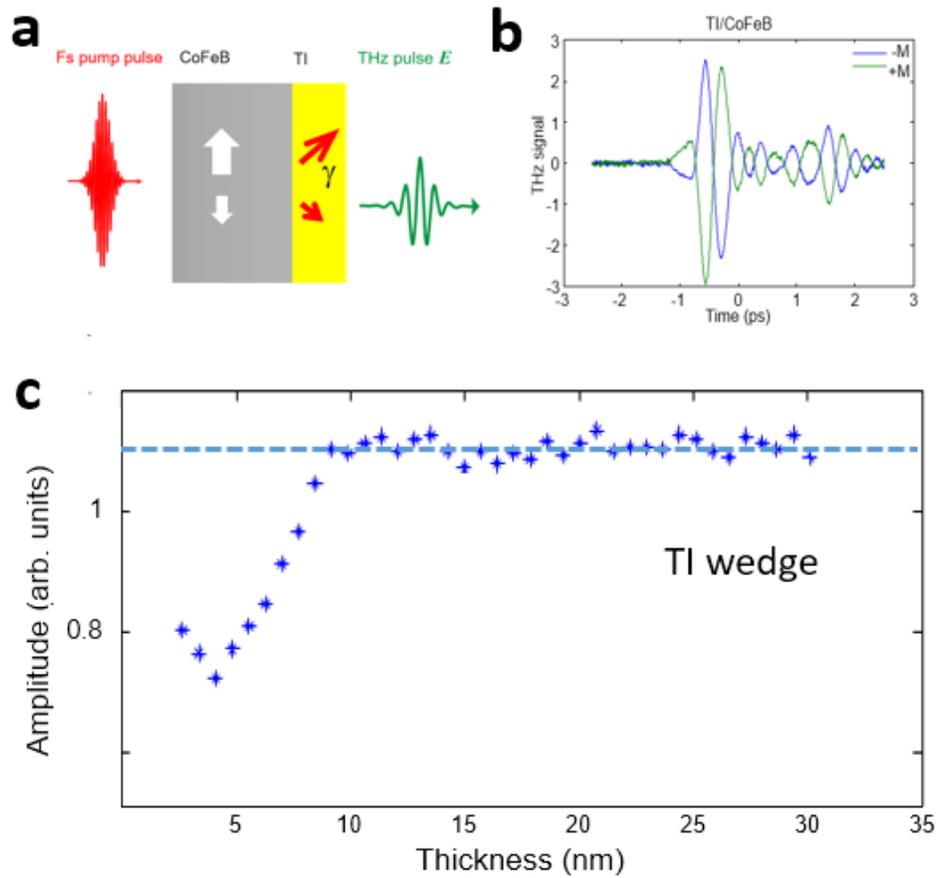

Figure 5: THz emission spectroscopy of ultrafast interfacial spin transport. a) Concept schematic. An ultrashort optical laser pulse induces spin transport between a ferromagnetic (CoFeB) and topological insulator (TI, $(Bi_{0.57}Sb_{0.43})_2Te_3$) thin film. Due to spin-to-charge current conversion, an ultrafast in-plane charge current results that emits a detectable THz electromagnetic pulse. b) Typical waveforms of THz electro-optic signals. They reverse almost completely when the in-plane magnetization $M$ of the FM layer is reversed. c) Amplitude of the THz emission signal as a function of the thickness of the TI layer.